\documentstyle[graphicx,pjw_cite]{mn}

\newcommand{\etal}{et~al.\ }
\def\la{\mathrel{\hbox{\rlap{\hbox{\lower4pt\hbox{$\sim$}}}\hbox{$<$}}}}
\def\ga{\mathrel{\hbox{\rlap{\hbox{\lower4pt\hbox{$\sim$}}}\hbox{$>$}}}}

\title[ROSAT observations of V471~Tau]{ROSAT observations of V471~Tauri, 
showing that stellar activity is determined by rotation, not age. }

\author[P.J. Wheatley]
	{Peter J. Wheatley \\
	 Department of Physics and Astronomy, University of Leicester, 
	     University Road, Leicester LE1 7RH }

\date{Accepted ... Received ...}

\begin{document}

\maketitle

\begin{abstract}
I present pointed ROSAT PSPC observations of the pre-cataclysmic 
binary V471~Tauri. 
The hard X-ray emission ($>$0.4\,keV) is not 
eclipsed by the K star, demonstrating conclusively 
that this
component cannot be emitted by the white dwarf. Instead I show that its
spectrum and luminosity are consistent with coronal emission from the 
tidally spun-up 
K star. 
The star is more active than other K stars in the Hyades, 
but equally active as K stars in the Pleiades with the same rotation periods,
demonstrating that rotation---and not age---is the key parameter in 
determining the level of stellar activity. 

Ths soft X-ray emission ($<$0.4\,keV) is emitted predominately by the white 
dwarf and is modulated on its spin period.
I find that the pulse-profile is stable on 
timescales of hours and years, supporting the idea that it is caused by
opacity of accreted material. The profile itself shows that the magnetic field
configuration of the white dwarf is dipolar and that the magnetic axis
passes through the centre of the star.

There is an absorption feature in the lightcurve of the white dwarf, which
occurs at a time when our line-of-sight passes within a stellar radius 
of the K star. The column density and duration of this feature imply a volume 
and mass for the absorber which are similiar to those of coronal mass 
ejections of the Sun. 

Finally I suggest that the spin-orbit beat period detected in the optical
by Clemens {\it et al.} may be the result of the interaction of the 
K-star wind
with the magnetic field of the white dwarf. 
\end{abstract}

\begin{keywords}
accretion, accretion discs --
binaries: close --
stars: activity --
stars: individual: V471 Tauri --
novae, cataclysmic variables --
white dwarfs --
X-rays: stars.
\end{keywords}

\section{Introduction}
The eclipsing close binary V471~Tauri is a member of the Hyades open cluster, 
and contains a 
white dwarf and 
K2V star in a 12.5\,h orbit \cite{Nelson70,Young72}. 
The white dwarf is hot, $\rm T=3\times10^{4}\,K$,
and is a strong source of 
ultraviolet and soft-X-ray emission \cite{Guinan84}. 
Observations with EXOSAT by \scite{Jensen86} revealed a double-peaked 
modulation of the soft X-rays at a period of 555\,s. 
They
suggested this 
could be caused by either the changing viewing angle of bright and dark 
regions on
the white dwarf, the 555\,s period being the 
rotation period of the white dwarf,
or radiatively-driven g-mode pulsations.
Bright or dark regions could be formed by magnetic
accretion of the K-star wind onto polar regions of the white dwarf: bright 
regions due to heating or dark regions due to opacity of 
accreted material.
The detection of the same period in the 
optical \cite{Robinson88,Clemens92}, but in anti-phase with the X-rays 
\cite{Barstow92}, proved that the modulation could not be due to 
pulsations. It also showed that the polar regions of the white dwarf are 
probably dark in X-rays and bright in the optical. 

\scite{Barstow92} presented ROSAT all-sky-survey observations of V471~Tau. They
detected the white-dwarf spectrum but also some X-ray emission at energies 
higher than could be attributed to the white dwarf ($>$0.3\,keV). 
Together with the observation that a fraction of the X-ray flux is not eclipsed
\cite{Jensen86}, this suggests that the K star is a significant source of 
X-rays.

In this paper I present observations of V471~Tau made with the ROSAT 
position-sensitive proportional counter (PSPC) during the pointed phase of the
mission. These confirm that the K star is a significant X-ray source, and show
that its emission is consistent with that expected from a rapidly-rotating 
K star. The pointed ROSAT observations are the first with sufficient 
sensitivity and spectral resolution to separate the X-ray emission 
of the white dwarf and K star. 

\section{Observations}
V471~Tau has been observed twice with the ROSAT PSPC 
\cite{Trumper83,Pfeffermann86} during the pointed phase 
of the mission: once in february 1991 with 4\,ks exposure, and once in 
august 1991 with 28\,ks exposure. The first observation was performed on-axis,
and so the lightcurves are modultated by occulations of the source by the PSPC
window support wires. 
The second was made with the standard 40\,arcmin offset, and is thus free 
from this contamination. In this paper I concentrate on the second 
observation.
It spans 24\,h, or roughly two orbital periods of V471~Tau, with a mean 
count rate $2.97\pm 0.01\rm\,s^{-1}$ (after corrections for vignetting 
and occultation by PSPC window support wires).
Spectra and lightcurves were
extracted from a region of 5.75\,arcmin radius around the source, and the 
background was 
estimated from a large region in another PSPC off-axis segment (choosen to be 
free of sources). Lightcurves were extracted in two energy bands, 
soft: pha\,8--39, 0.1-0.4\,keV; and hard: pha\,40-120, 0.4-1.2\,keV.

\section{Results}
\subsection{The eclipse and orbital-timescale variability}
\label{sec-555}
\begin{figure}
\begin{center}
\includegraphics{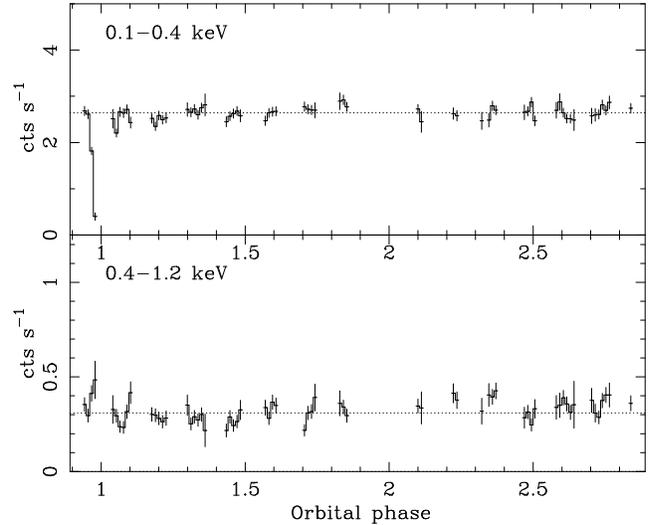}
\caption{\label{fig-555} ROSAT PSPC X-ray lightcurves of 
V471~Tau.
The lightcurves have been binned at the white-dwarf spin period 
(554.635\,s) and are plotted against orbital phase, where zero is the optical 
eclipse centre according to the ephemeris of Clemens \etal (1992). 
}
\end{center}
\end{figure}

The soft and hard ROSAT lightcurves of the August 1991 observation are plotted 
in Fig.\,\ref{fig-555}. 
They have been binned at the white-dwarf spin period 
(554.635\,s, \ncite{Barstow92}), and plotted against orbital phase using the 
ephemeris of \scite{Clemens92}. The observations span two orbital periods, and
one eclipse is covered partialy. 

\begin{figure}
\begin{center}
\includegraphics{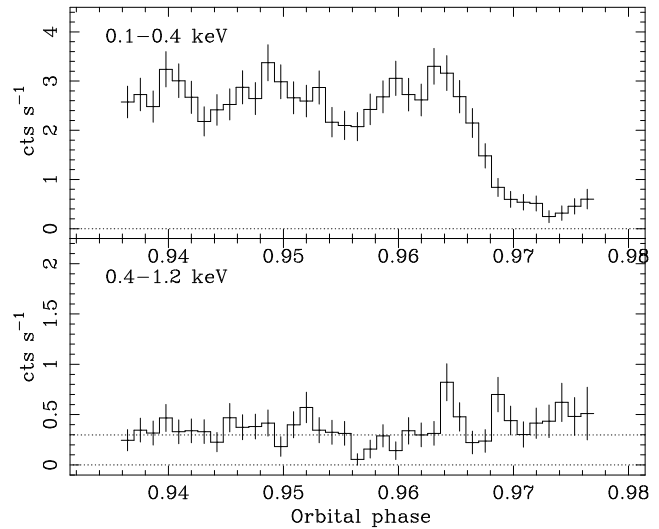}
\caption{\label{fig-50} The soft and hard lightcurves of the first observing 
slot, binned at 50\,s. The hard X-rays are not eclipsed. }
\end{center}
\end{figure}

Figure\,\ref{fig-50} shows the first data section  binned at 50\,s. 
The hard X-rays are not eclipsed, demonstrating conclusively that they are not 
emitted by the white dwarf. This effect was noted in the ROSAT all-sky survey 
data by \scite{Barstow92}, but at only marginal significance, due to the low 
exposure of individual survey scans. The K star is the obvious candidate 
source of the hard X-ray component (see Sect.\,\ref{sec-Kstar}). 

Outside eclipse, both lightcurves remain significantly variable. 
I find a reduced
$\chi^2$ of 1.6 with 56 degress of freedom (d.o.f.) for the soft lightcurve
when compared with its weighted mean count rate of $2.63\rm\, s^{-1}$.
And a $\chi^2$(d.o.f.) of 1.9(61) for the hard lightcurve compared with its
mean value of $0.31\rm\, s^{-1}$.
The variability of the soft lightcurve is dominated by a single bin in the 
second data slot. Removing this dip (see Sect.\,\ref{sec-dip} and 
Fig.\,\ref{fig-3slots}) the variability is barely significant, with 
reduced $\chi^2$ of 1.2(58). 
The RMS amplitude (excluding the eclipse and dip) 
is 0.13$\rm\,s^{-1}$ (5\%) for the soft 
and 0.06$\rm\,s^{-1}$ (19\%) for the hard lightcurve. 
Assuming the K star accounts for the variability in both bands, these 
amplitudes suggest it may emit up to a quarter of the soft-band flux.

The single narrow dip is the only 
sign of the 
deep orbital dips seen in a subset of EXOSAT observations \cite{Jensen86},
and an EUVE observation \cite{Cully96}.


\subsection{White-dwarf spin modulation}
\begin{figure}
\begin{center}
\includegraphics{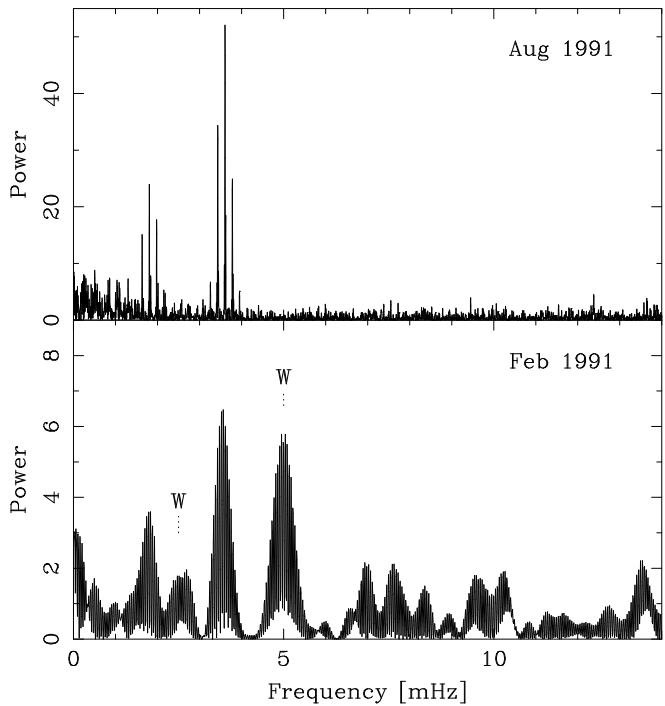}
\caption{\label{fig-ps} Power spectra of the soft X-ray ROSAT lightcurves of 
V471~Tau. The white-dwarf spin and its first harmonic stand out clearly.
There is no sign of the sideband modulation found at optical wavelengths. The
peaks labeled ``W'' in the lower panel are instrumental effects introduced
by the ``wobble'' of the ROSAT spacecraft. }
\end{center}
\end{figure}

The white-dwarf spin modulation, 
discovered by \scite{Jensen86} with EXOSAT, is
apparent in the soft X-ray lightcurve of Fig.\,\ref{fig-50}. 
In Fig.\,\ref{fig-ps} I show the
power spectra of the full soft X-ray lightcurves of both ROSAT PSPC 
observations. These were calculated using the Lomb-Scargle algorithm 
\cite{Scargle82}, as implemented by \scite{NumRec2}. 
Clearly the August 1991 spectrum is 
less heavily aliased
than that of the short
February 1991 observation, but the results are consistent. The two peaks marked
``W'' are introduced by the 400\,s ROSAT wobble, which is performed in order 
to blur the shadows of the PSPC window support wires. 
A periodic signal is detected with a period of 555\,s and an 
amplitude of 8\% (presumably the spin period), which has a strong 
first harmonic 
with 12\% amplitude, showing that the pulse profile is double peaked.
These results are identical to those found with 
EXOSAT by \scite{Jensen86}.
No periodic signal is found in the hard X-ray lightcurve. 

\scite{Clemens92} present optical observations of V471~Tau which are modulated
at the same periods, but also at the lower orbital sideband of the 555\,s 
period: 562\,s.
This peak is not apparent in our X-ray power spectra, 
and I find a 95\%-confidence upper limit to the amplitude of such a pulse of 
4\% (by fitting a sine function to the folded August 1991 lightcurve). 
In Sect.\,\ref{sec-conc-wd} I suggest this modulation may be a result of
the interaction of the wind of the K star and the magnetic field of the white 
dwarf.

\begin{figure}
\begin{center}
\includegraphics{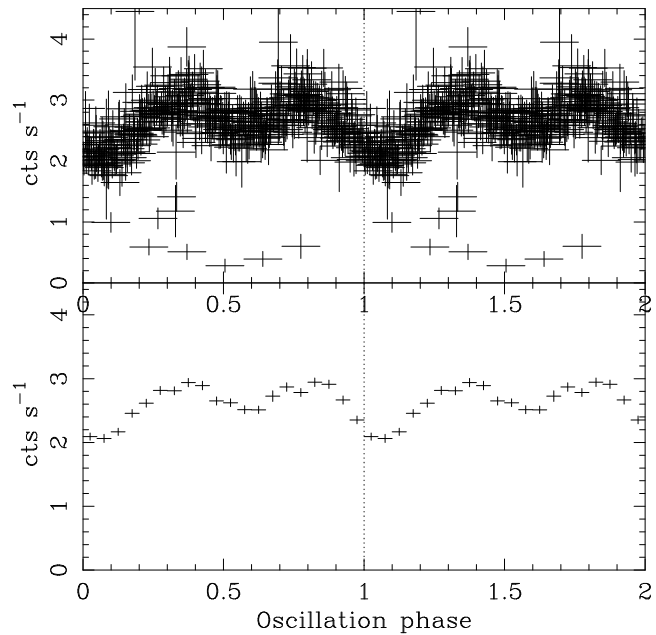}
\caption{\label{fig-fold} The soft-band PSPC lightcurve folded at the 
white-dwarf spin period: 554.635\,s. The top panel shows the individual 
data points, and the bottom shows them binned into twenty phase bins. 
}
\end{center}
\end{figure}

Figure\,\ref{fig-fold} shows the soft-band August 1991 lightcurve folded at 
the white-dwarf spin period. The top panel shows the individual data 
points positioned in phase (the low points are the eclipse and dip), 
and the bottom panel shows this binned into twenty phase bins. 
The RMS amplitude of the double-peaked pulse profile is 10\%. 
I have adopted 
the ephemeris of \scite{Barstow92}, for which the accumulated phase error is 
0.2. 
Within this error, the deep minimum is at the same phase as the minimum found 
by \scite{Barstow92} in the ROSAT WFC and PSPC all-sky survey data. 
This confirms their result that the X-ray and optical modulations are
in anti-phase. 
Barstow {\it et al.\ } found only a single minimum in their lightcurve, 
and suggested that 
the profile may have changed between the EXOSAT and ROSAT all-sky survey 
observations. 
The pointed ROSAT lightcurves, however, measured 
just 6\,months and 12\,months after the survey, 
have identical profiles to the EXOSAT observations. 
Given the poor 
phase coverage of the survey observations, and their poor statistical quality,
it seems most likely that the profile has remained constant throughout 
the period of EXOSAT and ROSAT observations.
Recent EUVE observations also show the same profile \cite{Dupuis97}. 

The top panel of Fig.\,\ref{fig-fold} shows that the scatter of individual 
points around the mean profile has similar amplitude to the statistical errors.
Comparison of data points (excluding eclipse and dip) with the mean profile 
yields a reduced $\chi^2$(d.o.f) of 1.06(1080).
Thus it seems the pulse profile is stable on short timescales (hours) as well
as the timescales between observations (years). 
Stability is important because the idea that the X-ray modulation is due to 
the opacity of accreted material at the magnetic poles of the white dwarf 
{\em requires} stability on timescales shorter than the diffusion timescale
of heavy elements in the white-dwarf atmosphere (estimated at 3\,yrs by 
\ncite{Vauclair79}). 

\subsection{The dip: coronal mass ejection?}
\label{sec-dip}
\begin{figure}
\begin{center}
\includegraphics{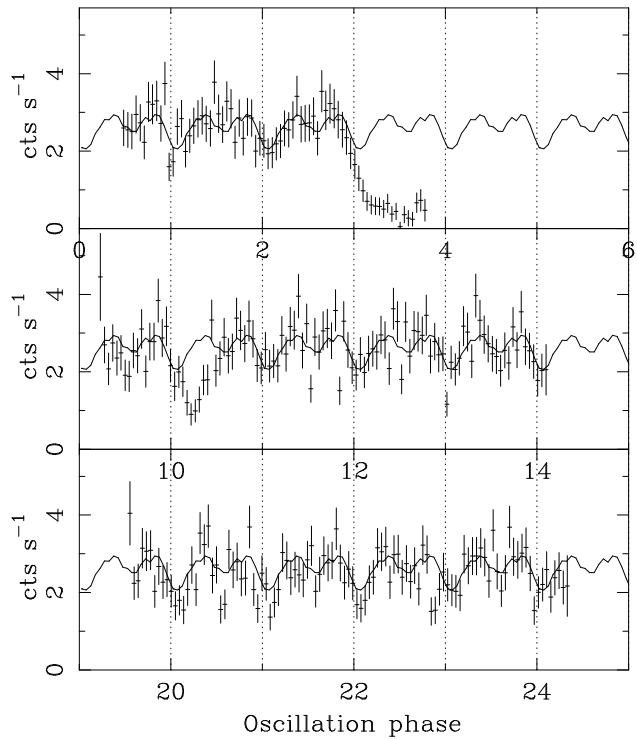}
\caption{\label{fig-3slots} 
Lightcurve of the first three slots in the PSPC observation of August 1991. 
The folded lightcurve of Fig.\,\ref{fig-fold} is overlayed. 
Apart from the eclipse, the only feature deviating from the simple two-peaked 
pulse profile in the entire
lightcurve is the dip at the beginning of the second slot 
(Sect.\,\ref{sec-dip}).   
}
\end{center}
\end{figure}

Figure\,\ref{fig-3slots} shows the first three sections of 
the August 1991 PSPC lightcurve, in which there are two obvious deviations
from the folded pulse profile of Fig.\,\ref{fig-fold} (which is overlayed in 
Fig.\,\ref{fig-3slots}). The first is the eclipse, which is discussed 
in Sect.\,\ref{sec-555}, and the second is the dip feature in the second 
slot. 
The third slot is typical of all the other
slots (not plotted). 

The dip 
is narrower than the dramatic orbital dips
seen in the first EXOSAT observation of V471~Tau \cite{Jensen86}. 
But it does occur
close to eclipse,
as did the EXOSAT dips and a 
similar dip observed 
with EUVE \cite{Cully96}. 
Of course at these times our line of sight passes close to the limb of the 
K star, and absorption by its wind and any magnetically supported structures 
may be expected to be most apparent. 
The dip occurs at orbital phase of 0.07, while the duration of the eclipse
is just 0.08. Thus the dip occurs when our line of sight passes within 
one stellar radius of the stellar surface.
The observations do not cover the dip phase a second time, so I can place no 
limits on its lifetime.
As may be expected, the dip is not apparent in the hard-band lightcurve, which
probably represents emission from the K star itself.

At mid dip the count rate falls to $0.9\rm\,s^{-1}$, from a mean of 
$2.63\rm\,s^{-1}$. From the eclipse light curve (Fig.\,\ref{fig-50}) I 
estimate the soft-band K star count rate is $0.4\rm\,s^{-1}$. Thus the 
white-dwarf count rate falls by a factor 4.5. The column density of neutral
material required to produce this drop, in this energy range, and with the 
best fitting spectrum of Sect.\,\ref{sec-Kstar} is 
$\rm N_H=9\times 10^{19}\,cm^{-2}$, with the usual assumptions of solar 
abundance and absorption cross sections given by \scite{Morrison83}.
The true column may be substantially greater than this because the absorber 
must be at least partially ionised.
Indeed Cully et al.\  find their dip is not apparent 
at long wavelengths suggesting that the
absorbing material must be highly ionised.
Both the EUVE and EXOSAT bandpasses extend to longer wavelengths than the 
ROSAT PSPC, so the greater widths of dips seen with those instruments may be 
due to their sensitivity to smaller column densities. 

One can estimate the size of the absorbing region from the duration of the
dip, $\sim$$230\rm\,s$. The relative velocity of the two stars in V471~Tau
is $\sim$$300\rm\,km\,s^{-1}$, so, assuming the absorber is close to and moves 
with the K star, it must have a width of $\sim$$4\times 10^{9}\rm\,cm$.
Assuming it has approximately the same depth along our line-of-sight as it
has width prepedicular to it, I find the absorber has a number density of 
Hydrogen atoms of $10^{10}\rm\,cm^{-3}$. 
Assuming further that it is spherical, the total mass of the absorbing medium 
must be of order $10^{16}\rm\,g$. 
These estimates of density and mass are similar to those measured for solar
coronal mass ejections \egcite{Wagner84,Kahler92}. 
So I suggest that the X-ray absorption
dips seen in V471~Tau are caused by material ejected from the K star in a 
similar manner.
\\

\subsection{The K star spectrum}
\label{sec-Kstar}
The ROSAT observations presented here are the first X-ray observations of 
V471~Tau in which the K2V star spectrum can be separated from that of the 
white dwarf. I make no attempt to investigate the spectrum of the white dwarf,
since this has been done using more appropriate data by \scite{Barstow97},
\scite{Dupuis97} and \scite{Werner97}. 
Instead I take the white-dwarf spectrum into account using the 
homogeneous hydrogen and helium model of \scite{Koester91}, 
and parameters derived from ORFEUS and IUE data by \scite{Barstow97}.  
I allowed only the normalisation and helium abundance to vary.
The K star spectrum was modelled
with one and two-temperature optically-thin 
thermal plasma models \cite{Mewe85,Kaastra93}, 
as the spectral response of the ROSAT 
PSPC does not merit a full differential emission measure treatment.

I applied these models to two spectra simultaneously. One extracted from 
eclipse, in which only the 
K star 
spectrum is visible (0.4\,ks), and one 
extracted from most of the remainder of the August 1991 observation 
(26\,ks).
All parameters were forced to be the same for the two data sets, except the
normalisation of the white-dwarf atmosphere model which was set to zero for
the eclipse spectrum. I also allowed the normalisation of the K star spectrum
to be different for the two data sets, but the relative importance of the 
two temperatures was forced to be the same. 

\begin{figure}
\begin{center}
\includegraphics{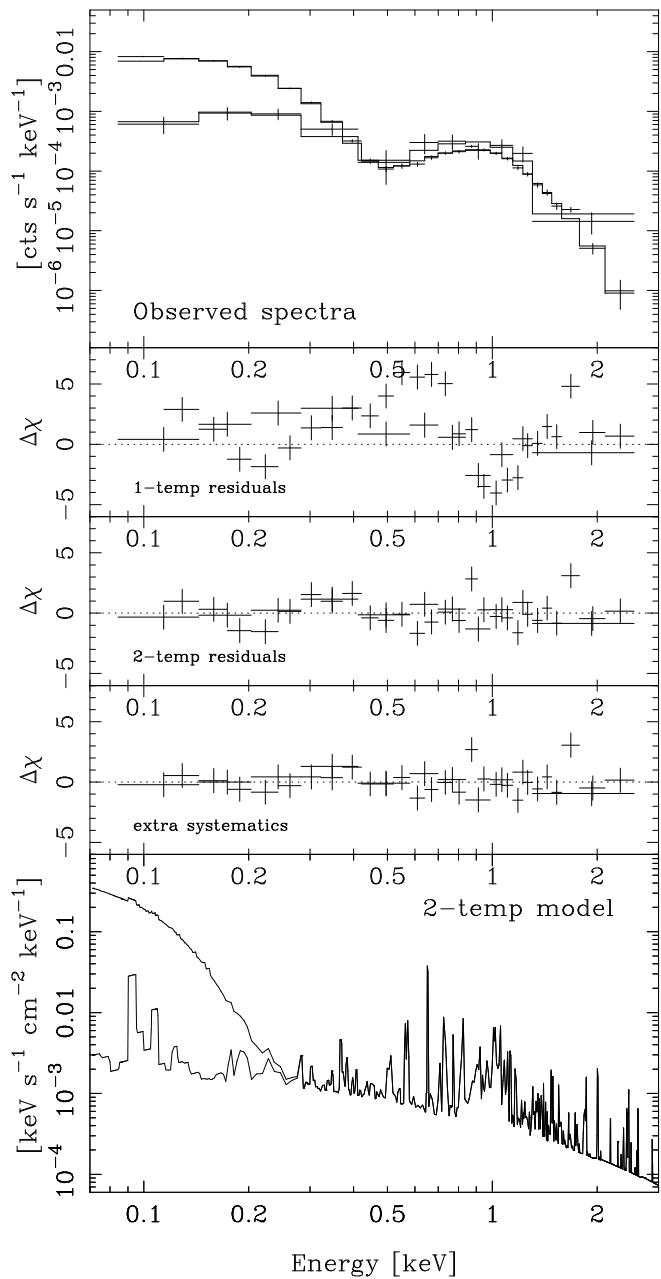}
\caption{\label{fig-spec} 
ROSAT PSPC spectra of V471~Tau. The top panel shows the raw counts spectra 
for eclipse and out-of-eclipse intervals. The next three panels show residuals
for different fits, discussed in the text. The bottom panel shows the model 
spectra for the best fit: a two-temperature fit to the K star spectrum, with 
the white-dwarf model fixed at that derived through fits to ORFEUS and IUE 
data.
}
\end{center}
\end{figure}

Figure\,\ref{fig-spec} shows the spectra and best-fitting model. 
The first bin in the out-of-eclipse spectrum shows strong positive residuals
in all my fits. This is probably due to the ``after pulse'' effect, 
introduced by contamination of the PSPC gas supply \egcite{Snowden94}.
I take the simplest approach to remove this effect by excluding this point 
from my fits. 
The second panel of Fig.\,\ref{fig-spec} shows the residuals of my 
single-temperature fit to the K star spectrum. This is an unacceptable fit with
a reduced $\chi^2$(d.o.f.) value of 8.7(32). 
The fit to the two-temperature model
(residuals in the third panel) is much better and marginally statistically 
acceptable with a reduced $\chi^2$ of 1.45(30). 
The strongest residuals occur below 0.4\,keV in the 
out-of-eclipse spectrum, and have the signature of the well-known temporal
gain variation problem of the PSPC \egcite{Prieto96}. 
I remove this effect by adding 
a somewhat arbitrary 10\% systematic error to the pha channels 8--40. This
results in an acceptable fit with reduced $\chi^2$ of 1.11(30)
(panel four of Fig.\,\ref{fig-spec}). I am satisfied that this is justified
especially because the out-of-eclipse spectrum
in this range
is dominated by the spectrum of the white dwarf---of which we are not 
concerned here. 
The two temperatures selected by this fit to the K star spectrum are 
$1.6\pm 0.2$\,keV and  
$0.36\pm 0.05$\,keV (68\%-confidence errors). The 0.1--2.5\,keV
flux corresponding to the normalisation of the (high-exposure) out-of-eclipse 
spectrum is $\rm (4.0\pm0.2)\times 10^{-12}\,erg\,s^{-1}\,cm^{-2}$. 
The interstellar column density is 
$\rm <7\times 10^{18}\,cm^{-2}$ (68\% confidence).
Using the Hipparcos distance of $47\pm4$\,pc \cite{Barstow97} yields a 
0.1--2.5\,keV luminosity of ($\rm 1.1\pm0.2)\times10^{30}\, erg\, s^{-1}$.
The bolometric correction is small, assuming of course that the fitted 
two-temperature model is an adequate description of the K star X-ray spectrum 
outside  the ROSAT band. I estimate that the total X-ray luminosity will be 
about 35\% larger than that in the range 0.1--2.5\,keV.

\subsubsection{Comparison with other active dK stars}
My results show that the uneclipsed X-ray emission of V471~Tau is somewhat 
different to that of other K stars in the Hyades.
\scite{Stern94} present one deep ROSAT 
pointing of part of the Hyades, and fit the spectra of a number of active 
K stars. They find, as I do, that single-temperature models do not 
adequately describe the spectra, but
that two-temperature models can. Their fitted temperatures, however, are
lower than those I find for V471~Tau. 
From their fits one would expect a K star to have fitted temperatures 
around 0.1--0.2\,keV and 0.5--1.0\,keV (rather than 0.4 and 1.6\,keV as I 
find). 

\scite{Pye94} present ROSAT data from a number of 
pointings of the 
Hyades. They do not attempt to fit spectra, but derive 
luminosity functions for dK and dM Hyads based on count rates.
They estimate an X-ray 
luminosity for each source by assuming a common distance (45\,pc) and assuming
that 1\,cts\,s$^{-1}$ in the PSPC corresponds to an X-ray flux of 
$6\times 10^{-12}\rm\, erg\,cm^{-2}\,s^{-1}$.
This method yields an estimate of $1\times10^{30}\rm\, erg\,s^{-1}$
for the luminosity of the uneclipsed source in V471~Tau 
(consistent with my measured value).
They detect thirteen of the seventeen K Hyads in their ROSAT fields, 
with X-ray luminosities in the range 
1.3--30$\rm\times 10^{28}\, erg\,s^{-1}$. Thus the most luminous K star 
Pye et al.\  find in the Hyades has a luminosity a factor three less than
I find is emitted by the K star in V471~Tau. 

\begin{figure}
\begin{center}
\includegraphics{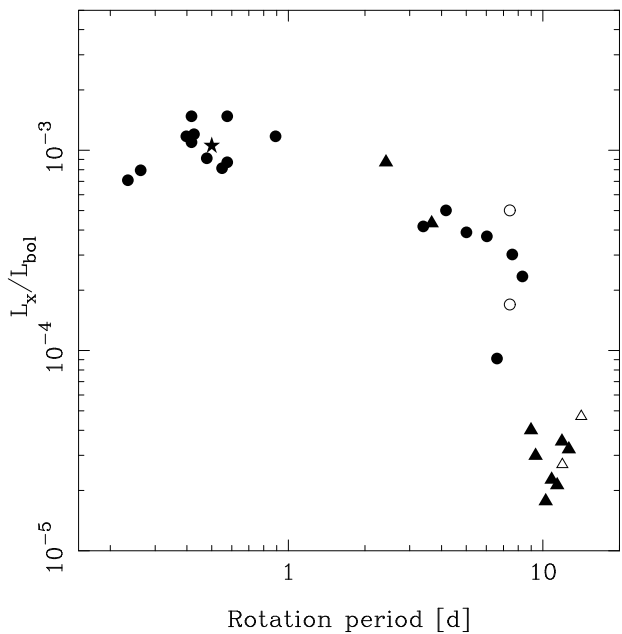}
\caption{\label{fig-xlum}
X-ray luminosity as a fraction of bolometric luminosity, plotted as a 
function of rotation period for a sample of Hyades K stars---triangles; 
Pleiades K stars---circles; 
and my measurement of the K star in V471~Tau---star.
Open symbols indicate upper limits to the X-ray luminosity. 
See text for references. }
\end{center}
\end{figure}

The most likely cause of the higher temperatures and excess X-ray luminosity 
is the high rotation rate imposed on the K star tidally 
by the 
white dwarf ($\rm P_{rot}=12.5\,h$). 
The correlation of magnetic activity and rotation rate
for late-type stars is well documented \egcite{Hartmann87}.
I can test this assertion by comparing my results with ROSAT observations
of the Pleiades \cite{Gagne95}, since the Pleiades is a younger open cluster
than the Hyades and its members rotate more rapidly.
Figure\,\ref{fig-xlum} shows very clearly that the rotation rate and X-ray 
luminosity of the K star in V471~Tau are indeed typical of K stars found in the
Pleiades. This figure shows the X-ray luminosity as a fraction of bolometric
luminosity (using the relation of \ncite{Johnson66}) for K stars 
($0.8<(B-V)<1.5$) in the Pleiades and Hyades as a function of their rotation
period. Data are taken from \scite{Pye94}, \scite{Stern95}, \scite{Gagne95},
\scite{Radick87} and \scite{Prosser95}.
Remarkably, the K Pleiads and Hyads seem to obey the same activity-rotation
relation. To the best of my knowledge, this comparison has not been made 
before. 
The bolometric luminosity of the K star in V471~Tau is calculated from its
spectral type K2V, rather than its B--V colour, since the B magnitude is
contaminated by the white dwarf.
Its presence 
amoung the Pleiads shows that the
X-ray emission of late-type stars depends on rotation only, and not on their
age. 
Finally, comparison with the spectral fits of \scite{Gagne95} shows that my 
fitted X-ray temperatures are also typical of K Pleiads. 
I thus conclude that the 
uneclipsed X-ray emission of V471~Tau is indeed consistent with that expected 
for a rapidly rotating K star.

\section{Conclusions}
\subsection{The K star}
I show conclusively, for the first time, that the hard X-ray component of 
V471~Tau is not eclipsed by the K star. I find that the spectrum and 
luminosity of this component is typical of rapidly-rotating K stars, and
conclude that it is most-likely emitted by the K star. 

I note that this X-ray flux is normal for its rotation rate rather than its
age (Fig.\,\ref{fig-xlum}). This demonstrates that rotation is the key 
parameter in determining stellar activity, and not age. 
In most studies, age and rotation rate are too closely related to be separable.

I also observe an absorption dip in the X-ray emission of the white dwarf, 
absent in the K-star flux, which implies a column density and length-scale 
for the absorber similar to coronal mass ejection events seen on the Sun. 
I conclude that similar processes are probably at work on the K star in 
V471~Tau, and may account for the more dramatic orbital dips seen with 
EXOSAT and EUVE (because both these instruments are sensitive to a smaller 
absorbing column density).

\subsection{The white dwarf}
\label{sec-conc-wd}
I detect the 555\,s X-ray modulation discovered by \scite{Jensen86}. 
The pulse-profile is identical to that measured with EXOSAT and EUVE, and is
also stable throughout the ROSAT observations. 
This is consistent with the idea that 
the modulation is caused by opacity of material accreted at the magnetic poles
of the white dwarf (since this opacity cannot vary on timescales shorter than 
the diffusion of metals in the atmosphere of the white dwarf). 

The profile itself is double-peaked, and the two peaks are separated in phase
by precisely 180$^\circ$. If the modulation is indeed caused by the opacity of
accreted material, then the magnetic field of the white dwarf must direct the 
accretion flow onto two regions directly opposing each other. This shows that
the magnetic field configuration must be dipolar, and that the magnetic axis 
must pass though the centre of the white dwarf. 

I do not detect the beat pulse discovered in the optical by \scite{Clemens92}
and place a 95\% upper limit to its X-ray amplitude of 4\%.
Clemens {\it et al.\  } interpret this pulse as reprocessing of ionising 
radiation
on the face of the K star. However the beat pulse is single
peaked while the X-ray pulse is double peaked, and Clemens {\it et al.\  }
can only reconcile these facts by assuming the X-ray emission does not trace 
the bulk of the ionising radiation. 
Instead I suggest that the beat pulse may reflect a modulation in accretion
rate caused by the interaction of the wind of the K star with the magnetic 
field of the white dwarf. Beat periods arise naturally in 
systems where the accretion flow 
has memory of 
orbital phase \egcite{Warner86}. 
However the same problem arises with this interpretation
because in a high-inclination system, such as V471~Tau, where we see both
magnetic poles, one would expect to see modulation at the frequency 
$\rm (2\,f_{spin}-f_{orbit})$ rather than that detected by 
Clemens {\it et al.\  } $\rm (f_{spin}-f_{orbit})$ \cite{Wynn92}. 
Still, as the geometry of magnetic wind accretion is poorly known, there may be
scope to explain the single peaked beat pulse.

\section{Acknowledgements}
PJW is a postdoctoral fellow of PPARC. ROSAT data were extracted from the 
Leicester data archive at Leicester University.

\bibliographystyle{$papers/tex/bibtex/mnras}
\bibliography{$papers/tex/bibtex/mn_abbrev,$papers/tex/bibtex/refs}

\end{document}